\documentclass[a4paper,12pt]{article}
\usepackage{verbatim} 
\usepackage{jheppub} 

\usepackage[T1]{fontenc} 
\usepackage{amsmath}
\usepackage[percent]{overpic}
\usepackage{wrapfig}
\usepackage{tabu}
\usepackage{diagbox}

\usepackage{graphicx}
\usepackage{tikz}
\usepackage{import}
\usepackage{accents}
\usepackage{mathrsfs,amsmath,amssymb,slashed}
\usepackage{multirow,multicol}
\usepackage[percent]{overpic}
\usepackage{slashed}

\usepackage{wrapfig}
\usepackage{tabu}
\usepackage{diagbox}
\usepackage{mathrsfs,amsmath,amssymb,amsthm,amsfonts,tikz,graphicx,accents,hyperref,color}
\usepackage{leftidx}
\usepackage{import}
\usetikzlibrary{decorations.pathmorphing}
\DeclareFontFamily{OT1}{rsfs}{}

\DeclareFontShape{OT1}{rsfs}{m}{n}{ <-7> rsfs5 <7-10> rsfs7 <10->rsfs10}{} 

\DeclareMathAlphabet{\mycal}{OT1}{rsfs}{m}{n}

\newcommand{\nn}{\nonumber}

\newcommand{\be}{\begin{equation}}
\newcommand{\ee}{\end{equation}}

\DeclareMathOperator{\extdm}{d}
\newcommand{\extd}{\extdm \!}

 \title{\bf 
  {Warped Schwarzian theory}}
\author[]{Hamid R. Afshar}
\affiliation[]{\it Institute for Theoretical Physics, TU Wien, Wiedner Hauptstr.~8, A-1040 Vienna, Austria}
\affiliation[]{\it School of Physics, Institute for Research in Fundamental
Sciences (IPM),\\ P.O.Box 19395-5531, Tehran, Iran}
 
 \emailAdd{afshar@ipm.ir}

\abstract{
We consider the (twisted) warped Virasoro group Diff($S^1$)$\ltimes$ C$^\infty$($S^1$) in the presence of its three cocycles. We compute the Kirillov-Kostant-Souriau symplectic 2-form on coadjoint orbits. We then construct the Euclidean action of the `warped Schwarzian theory' associated to the orbit with SL(2,$\mathbb{R}$)$\times$U(1) stabilizer as the effective theory of the reparametrization over the base circle and evaluate the corresponding one-loop-exact path integral. We further discuss thermodynamics of the wSch theory in comparison with the complex SYK model. 

}
\begin{document}
\maketitle

\section{Introduction}
The Sachdev-Ye-Kitaev (SYK)  model \cite{Sachdev:1992fk,Kitaev,Kitaev:2017awl} has attracted
much attention recently as a toy model for AdS/CFT in lowest possible dimensions. It is an interacting  statistical quantum mechanical model of $N$ Majorana fermions with random couplings. In the regime of large coupling (low temperature) this model is perturbatively (in the $1/N$-expansion) dual to a simple theory of two-dimensional dilaton-gravity known as the Jackiw-Teitelboim (JT) gravity \cite{Teitelboim:1983ux,Jackiw:1984je}.\footnote{A variant of the SYK model is the the so-called complex SYK model \cite{Sachdev:2015efa,Davison:2016ngz} in which fermions are complex instead of being Majorana.}

A universal feature emerging for all variety of these quantum mechanical models is their low energy dynamics which is governed by  a solvable Euclidean theory known as  the `Schwarzian theory' \cite{Maldacena:2016hyu} which turns out to be the appropriate boundary term defining our two dimensional dilaton-gravity theory. The dynamical field in this theory is a quasi-periodic field $f(\tau)$ with $\tau$  being the imaginary time on the thermal circle $0<\tau<\beta$ and performing as the reparametrization on the thermal circle Diff($S^1$). This effective theory is by construction defined on the symplectic manifold Diff($S^1$)/SL(2,${\mathbb R}$) or Diff($S^1$)/U(1) and its quantum mechanical realization has been shown to be one-loop exact \cite{Stanford:2017thb}. 
These symplectic manifolds are technically known as coadjoint orbits of the Virasoro group \cite{Witten:1987ty,Alekseev:1988ce,Bakas:1988mq,Rai:1989js}.  The coadjoint orbit method \cite{kirillov,kostant,Souriau,Wiegmann:1989hn}  which we will benefit from in this work gives a geometric interpretation of the coadjoint representation of Lie groups and provides a systematic to construct field theory actions on a given Lie group orbit. For reviews and other applications and employment of the orbit method see \cite{Cotler:2018zff,Mandal:2017thl,math-ph/0602009,Barnich:2017jgw,10.1007/978-1-4612-5547-5_5,Penna:2018xqq,Oblak:2016eij,Barnich:2014kra,Delius:1990pt}.

 One of the consequences of the exactness of the path integral of the Schwarzian theory is that the Hamiltonian function associated to the imaginary time evolution is in fact proportional to the Schwarzian action itself.
In this paper we use this consequence as a base for constructing the `warped Schwarzian theory' in section \ref{WSCHT} based upon the coadjoint orbits of the warped Virasoro group \cite{Afshar:2015wjm} which is reviewed and developed for our application  in section \ref{CoadjointWVG}. We also calculate the Kirillov-Kostant-Souriau symplectic 2-form \cite{kirillov,kostant,Souriau,Wiegmann:1989hn} on the corresponding coadjoint orbits in section \ref{symplecticformWVG} and then use it as a measure to evaluate the one-loop-exact path-integral of the theory in section \ref{pathintegral} and discuss the spectrum of the theory in terms of the density of states. In section \ref{discussion} we discuss the thermodynamics of the warped Schwarzian theory and its implication to the complex SYK model.

\section{Warped Virasoro symmetry}\label{CoadjointWVG}

The Virasoro group is the  unique nontrivial central extension of the group of reparametrization on the circle. It has several generalizations but a simple extension is to make a semi-direct product of it with functions on the circle. We denote it as `warped Virasoro group', a terminology inferred from the physics literature \cite{Hofman:2011zj,Detournay:2012pc} and mostly studied in the context of two dimensional  warped conformal field theories (WCFTs).\footnote{In mathematics literature it is classified as the Schr\"odinger-Virasoro Lie group \cite{Roger:2006rz} or the corresponding algebra is sometimes called the Heisenberg-Virasoro algebra  \cite{arbarello1988,billig2003}.} Here in this work we are interested in the one dimensional (Euclidean time) realization of this symmetry group. Recently 
it has been shown that the warped Virasoro algebra is the underling symmetry for the low energy effective action of the complex SYK model
\cite{Chaturvedi:2018uov}.

\subsection{Coadjoint orbits of the warped Virasoro group}\label{coadjointsec}

The warped Virasoro group $\hat{G}$ is the universal central extension of the Lie group $G$ of all orientation preserving diffeomorphism of the circle acting naturally on smooth functions of the circle;
\begin{align}\label{TWVgroup}
    (f,\sigma) \in G\equiv\text{Diff}(S^1)\ltimes \text{C}^\infty(S^1)\,.
\end{align} 
For $f\in\text{Diff}(S^1)$ and $\sigma\in C^\infty(S^1)$ the group operation is
\begin{align}\label{groupoper}
(f_1,\sigma_1)\cdot(f_2,\sigma_2)=(f_1\circ f_2\,,\,\sigma_1+\sigma_2\circ f_1^{-1})    \,,
\end{align}
where the circle $\circ$ denotes the function composition. This group operation is inferred from the semi-direct product structure of the group \cite{Barnich:2015uva,Afshar:2015wjm}.

The corresponding Lie algebra $\hat{\mathbf g}$, called the warped Virasoro algebra is defined as the unique (up to isomorphism) nontrivial central extension of the Lie algebra ${\mathbf g}$ of first order differential operators on $S^1$;
\begin{align}\label{adjvector}
\mathbf{v}(\tau)\equiv\epsilon(\tau)\partial_\tau +\sigma(\tau)\,\in {\mathbf g}\,.
\end{align}
with $\epsilon(\tau+\beta)=\epsilon(\tau)$ and $\sigma(\tau+\beta)=\sigma(\beta)$ where $\beta$ is the circumference of $S^1$. This Lie algebra admits three central extensions \cite{marshall1990,Ovsienko:1994im,Roger:2006rz,Unterberger:2011yya}. We denote each element of $\hat{\mathbf g}={\mathbf g}\oplus{\mathbb R}^3$ by a pair $(\mathbf{v}(\tau),\mathbf{a})$ with $\mathbf{a}=a_1\mathbf{e}_1+a_2\mathbf{e}_2+a_3\mathbf{e}_3\in{\mathbb R}^3$ where $\mathbf{e}_i$ can be considered as orthonormal basis of ${\mathbb R}^3$. The commutator in the centrally extended Lie algebra $\hat{\mathbf g}$ is defined as
\begin{align}
    \big[(\mathbf{v}_1(\tau),\mathbf{a}_1)\,,(\mathbf{v}_2(\tau),\mathbf{a}_2)\big]=\big([\mathbf{v}_1,\mathbf{v}_2](\tau),\eta(\mathbf{v}_1,\mathbf{v}_2)\big)\,,
\end{align}
with
\begin{align}
&[\mathbf{v}_1,\mathbf{v}_2](\tau)=\big(\epsilon_1(\tau)\epsilon_2'(\tau)-\epsilon_1'(\tau)\epsilon_2(\tau)\big)\partial_\tau+\big(\epsilon_1(\tau)\sigma_2'(\tau)-\epsilon_2(\tau)\sigma_1'(\tau)\big)\,,\\
&\eta(\mathbf{v}_1,\mathbf{v}_2)=\int_{S^1}\left[\tfrac{1}{12}\big(\epsilon_1'(\tau)\epsilon_2''(\tau)\big)\mathbf{e}_1+\tfrac{1}{4}   \big(\sigma_2(\tau)\epsilon_1''(\tau)-\sigma_1(\tau)\epsilon_2''(\tau)\big) \mathbf{e}_2 +\tfrac{1}{2}\big(\sigma_1'(\tau)\sigma_2(\tau)\mathbf{e}_3 \big)\right],
\end{align}
where we have 
introduced the notation $\int_{S^1}\equiv\int\limits_0^\beta\extd\tau$. Note that at this point $\beta$ is just a parameter specifying the size of the circle $S^1$, later in our physical setup it will denote the inverse temperature of our quantum mechanical system on $S^1$.

Corresponding to the adjoint vectors $\mathbf{v}(\tau)\in {\mathbf g}$, we have coadjoint vectors $\mathbf{b}(\tau)\in {\mathbf g}^*$ which maps adjoint vectors to numbers. In the case of the warped Virasoro algebra these covectors can be represented in terms of a quadratic density and a one-form on the circle;
\begin{align}\label{covector}
\mathbf{b}(\tau)\equiv T(\tau)\extd\tau^2 +P(\tau)\extd\tau\,\in {\mathbf g}^*\,.
\end{align} A coadjoint vector of the warped Virasoro algebra $\hat{\mathbf g}$ is a pair $(\mathbf{b}(\tau),\mathbf{c})\in\hat{\mathbf g}^*$  where $\mathbf{c}=c_1\mathbf{e}_1^*+c_2\mathbf{e}_2^*+c_3\mathbf{e}_3^*$ is the dual of $\mathbf{a}$ such that $\langle \mathbf{e}_i^*,\mathbf{e}_j\rangle=\delta_{ij}$. The pairing between $\hat{\mathbf g}\equiv{\mathbf g}\oplus{\mathbb R}^3$ and $\hat{\mathbf g}^*\equiv{\mathbf g}^*\oplus{\mathbb R}^3$ is defined by
\begin{align}\label{pairing}
    \langle(\mathbf{b},\mathbf{c})\,,(\mathbf{v},\mathbf{a})\rangle=\int_{S^1}\big(T(\tau)\epsilon(\tau)+P(\tau)\sigma(\tau)\big)+c_1 a_1+c_2 a_2+c_3 a_3\,.
\end{align}
The group ${\hat G}$ acts on vectors by the adjoint action Ad$_{(f,\sigma)}$ and on covectors by the coadjoint action Ad$_{(f,\sigma)}^*$. The pairing \eqref{pairing} should be invariant under the group action
so the coadjoint representation Ad$^*$ of the group $\hat G$ is defined by \footnote{The central elements spanning ${\mathbb R}^3$ in the group ${\hat G}\equiv G\times {\mathbb R}^3$ acts trivially on the (co)algebra and so the (co)adjoint action of ${\hat G}$ is the same as of $G$. For detailed analysis see appendix A of \cite{Afshar:2015wjm}.}
\begin{align}
    \langle \text {Ad}_{(f,\sigma)}^*(\mathbf{b},\mathbf{c}),(\mathbf{v},\mathbf{a})\rangle=\langle(\mathbf{b},\mathbf{c}),\text {Ad}_{(f,\sigma)}^{-1}(\mathbf{v},\mathbf{a})\rangle\,.
\end{align}
 The coadjoint action leaves the central elements invariant but transforms the $\mathbf{b}(\tau)$ covector into a new covector in the coadjoint orbit
 \begin{align}\label{coadjrep}
     \text{Ad}_{(f,\sigma)^{-1}}^*(\mathbf{b},\mathbf{c})=(\mathbf{b}\circ f-\mathbf{c}\cdot \mathbf{S}(f,\sigma),\mathbf{c})\,,
 \end{align}
 where 
\begin{align}
     \mathbf{b}\circ f={T}(f(\tau))\extd f^2-P(f(\tau))\extd\sigma\extd f+{P}(f(\tau))\extd f\,,
 \end{align}
 is the natural action of the group $G=$Diff($S^1$)$\ltimes$C$^\infty$($S^1$) on the dual space of its Lie algebra $\mathbf{g}^*$ while $\mathbf{S}(f,\sigma)$ term can be considered as the contribution of the central extension of the group;
 \begin{align}
     \mathbf{S}(f,\sigma)=\bigg(\mathbf{e}_1\left[\big(\tfrac{f''}{f'}\big)'-\tfrac12\big(\tfrac{f''}{f'}\big)^2\right]+\mathbf{e}_2\big(\sigma''-\sigma'\tfrac{f''}{f'}\big)-\mathbf{e}_3\tfrac{\sigma'^2}{2}\bigg)\extd\tau^2+\left(\mathbf{e}_3\sigma'+\mathbf{e}_2\tfrac{f''}{f'}\right)\extd\tau\,.
 \end{align}
 \subsection{Vacuum orbit on the cylinder}
Each coadjoint orbit is parametrized by $f$ and $\sigma$ and identified uniquely w.r.t. an orbit representative ($\mathbf{b}_0,\mathbf{c}$) where
\begin{align}\mathbf{b}_0=T^{\text{\scriptsize vac}}\extd\tau^2+P^{\text{\scriptsize vac}}\extd\tau\,,\end{align}
 In order to determine the orbit representative we visualize the warped Virasoro group as the following infinitesimal transformation on the $\mathbb{R}^2$ plane with coordinates $(\tau, x)$;
\begin{align}\label{infinittrans}
\tau\to \tau +\epsilon(\tau)\,,\qquad\text{and}\qquad x\to x +\sigma(\tau)\,.
\end{align}
Upon integrating the infinitesimal
transformations \eqref{infinittrans}, one obtains finite diffeomorphisms of the plane $\mathbb{R}^2$
given by \footnote{The coordinate $\tau$ does not have to be periodic in general.}

\begin{align}\label{groupaction}
\tau\to f(\tau)\qquad\text{and}\qquad x\to x +g(\tau)\,,
\end{align}
where $g(\tau)=\sigma\circ f(\tau)$. The reparametrization $f(\tau)$ and the translation $g(\tau)$ are respectively an orientation preserving conformal transformation and an arbitrary function on a line forming the group $\text{Diff}(\mathbb{R})\ltimes \text{C}^\infty(\mathbb{R})$. In our physics setup,  $\tau$ is parametrizing the temperature of the system whereas the pure imaginary line $x$  (not to be confused by any spacetime coordinate) parametrizes a direction along a new chemical potential $\mu$ as we will see later on. 

Associated to these group elements, the pair of currents $(T(\tau),P(\tau))$ generate infinitesimal general coordinate transformations along Euclidean time $\tau$  and translation along $x$ as in \eqref{infinittrans} respectively. This group admits three central extensions $(c,\kappa,k)$, therefore these currents are anomalous which can be observed in their transformation under the coadjoint representation of $G$  as we briefly reviewed above and more elaborated in \cite{Afshar:2015wjm}.

In order to consider the theory at finite temperature $T=\beta^{-1}$ we replace $\mathbb{R}$ by 
$S^1$ with $\tau\sim\tau+\beta$. In this case $g(\tau)$ is an arbitrary function on the circle and $f(\tau)$ is a
diffeomorphism of the circle; 
\begin{align}\label{boundarycond}
    f(\tau + \beta) = f(\tau) + \beta\,,\qquad\quad g(\tau + \beta) = g(\tau)\,.
\end{align}
Relative to the line, the vacuum energy on the circle is shifted. We require the vacuum on the circle to be SL(2,$\mathbb R$)$\times$U(1) invariant so we can determine these vacuum state values  which correspond to the orbit representative uniquely
by mapping between the cylinder  and the plane using
the following set of finite warped conformal transformation;
\begin{align}\label{cylintoplain}
    \tau\to e^{2\pi i\tau/\beta}\,,\qquad x\to x+\alpha\,\tau\,.
\end{align}
where $\alpha$ is an arbitrary pure imaginary tilt \cite{Detournay:2012pc} quantifying the chemical potential $\mu=i\alpha$ in the WCFT context \cite{Song:2017czq}. We have the following thermal identification;
\begin{align}
    (\tau,x)\sim (\tau+\beta, x-i\mu\beta)\,.
\end{align}
In our one dimensional (quantum) mechanical system one should not consider $x$ as a real space coordinate but as the coordinate on the phase space parametrizing a new chemical potential $\mu$.

The coadjoint representation \eqref{coadjrep} provide the finite transformation law of the currents $T$ and $P$ under a finite
transformation \eqref{groupaction}. These transformation rules map the pair ($T,P$) to a new pair ($\tilde T,\tilde P$) at $f(\tau)$ whose explicit form can be read from \eqref{coadjrep} by setting $c_1=c$, $c_2=4\kappa$ and $c_3=-k$  according to the notation in  \cite{Afshar:2015wjm};
\begin{subequations}\label{trans2}
\begin{align}
\tilde{P}(f(\tau))&=\frac{1}{f'(\tau)}\left[P(\tau)+\kappa\frac{f''(\tau)}{f'(\tau)}-\frac{k}{2}g'(\tau)\right]\\
\tilde{T}(f(\tau))&=\frac{1}{f'(\tau)^2}\bigg[T(\tau)+\frac{c}{12}\{f(\tau);\tau\}-P(\tau)g'(\tau)-\kappa g''(\tau)+\frac{k}{4}g'(\tau)^2\bigg]\,.
\end{align}
\end{subequations}
Here $g=\sigma\circ f$ and $\{f;\tau\}=\left(\frac{f''}{f'}\right)'-\frac12\left(\frac{f''}{f'}\right)^2$ denotes the Schwarzian derivative  and prime refers to derivative w.r.t. $\tau$.
The infinitesimal form of these transformation can be obtained by expanding $f$ and $g$ to linear order of $\epsilon$ and $\sigma$;
\begin{subequations}\label{infinittranstran}
\begin{align}
    \delta_{(\epsilon,\sigma)} P &=\epsilon P'+\epsilon'P-\kappa\epsilon''+\frac{k}{2}\sigma'\,,\\
    \delta_{(\epsilon,\sigma)} T &=\epsilon T'+2\epsilon'T-\frac{c}{12}\epsilon'''+\sigma'P+\kappa\sigma''\,.
\end{align}
\end{subequations}

Plugging $(f(\tau),g(\tau))=(e^{2\pi i\tau/\beta},\alpha\,\tau)$ into \eqref{trans2} and setting the vacuum values on the plane to zero 
we obtain the vacuum values on the cylinder;
\begin{align}\label{vacval}
    P^{\text{\scriptsize vac}}=-\frac{2\pi i\kappa}{\beta}+\frac{\alpha\,k}{2}\,,\qquad T^{\text{\scriptsize vac}}=-\frac{\pi^2c}{6\beta^2}-\frac{2\pi i\kappa\,\alpha}{\beta}+\frac{\alpha^2k}{4}\,.
\end{align}
If we choose the exponential map by $f^{-1}=e^{-2\pi i\tau/\beta}$ we are led to the same expression as in \eqref{vacval} with $\beta\to-\beta$. This means that the $\mathbb{Z}_2$-symmetry of interchanging between degenerate vacua $|\beta\rangle$ and $|-\beta\rangle$ is broken by the presence of the $\kappa$-term. This map amounts to choosing only one of these vacuum on the cylinder and as a consequence our final result will be sensitive to this. This symmetry is however restored if we  simultaneously change $\kappa\to-\kappa$.
We have chosen $g(\tau)$ in \eqref{cylintoplain} to be linear in $\tau$ such that the $(T,P)$ are transformed only by constant values which amounts to a shift of only zero modes. It is important to note that for having a well-defined quantum theory defined on the circle we have $\alpha$ and $g(\tau)$ being pure imaginary.  Otherwise the Energy  ground state value $T^{\text{\scriptsize vac}}$ will have an imaginary part on the circle. This is consistent with the holographic constraints on WCFTs \cite{Detournay:2012pc,Apolo:2018eky}. 

By assigning vacuum values \eqref{vacval} to $(\tilde{T},\tilde{P})$ in \eqref{trans2} we generate all vacuum-representatives of the SL(2,$\mathbb R$)$\times$U(1) invariant warped Virasoro coadjoint orbit on the cylinder;
\begin{subequations}\label{constrep}
\begin{align}
P(\tau)&=\frac{k}{2}\left(g'+\alpha\,f'\right)-\kappa\left(\frac{f''}{f'}+\frac{2\pi i}{\beta}f'\right)\\
T(\tau)&=-\frac{c}{12}\left\{\tan\tfrac{\pi }{\beta}f;\tau\right\}+\frac{k}{4}(g'+\alpha f')^2-\kappa\bigg[\frac{2\pi i\alpha }{\beta}f'^2+\left(\frac{f''}{f'}+\frac{2\pi i}{\beta}f'\right)g'-g''\bigg] \,,
\end{align}
\end{subequations}
with the property $\big\{\tan\tfrac{\pi }{\beta}f;\tau\big\}=\{f;\tau\}+\frac{2\pi^2}{\beta^2}f'^2$. 
Subject to boundary conditions \eqref{boundarycond} on fields $f(\tau)$ and $g(\tau)$ the Fourier mode  generators of \eqref{constrep} on the cylinder
\begin{align}
    L_n=\int\limits_0^\beta\extd\tau \,[\underbrace{T(\tau)-T^{\text{\scriptsize vac}}}_{T^{cyl}(\tau)}]\,e^{\frac{2\pi}{\beta} in\tau }\,,\qquad\quad P_n=\int\limits_0^\beta\extd\tau\,[ \underbrace{P(\tau)-P^{\text{\scriptsize vac}}}_{P^{cyl}(\tau)}]\,e^{\frac{2\pi}{\beta} in\tau }\,,
\end{align}
satisfy the twisted warped Virasoro algebra;
\begin{subequations}\label{wcftalg}
\begin{align}
i\{L_n,L_m\}&=(n-m)L_{n+m}+\frac{c}{12}(n^3-n)\delta_{n+m,0}\,,\\
i\{L_n,P_m\}&=-mP_{n+m}-i\kappa(n^2-n)\delta_{n+m,0}\,,\\
i\{P_n,P_m\}&=\frac{k}{2}n\,\delta_{n+m,0}\,,
\end{align}
\end{subequations}
where we used \eqref{infinittranstran} with $\epsilon(\tau)\sim e^{\frac{2\pi}{\beta} in\tau }$ and $\sigma(\tau)\sim e^{\frac{2\pi}{\beta} in\tau }$, and the fact that $\{Q_1,Q_2\}=-\delta_1Q_2$. 
It is clear form the algebra \eqref{wcftalg} that the global part of this symmetry algebra is four dimensional corresponding to the invariance of the vacuum under  ($L_{\pm1,0}, P_0$)  or under ($L_{1,0}, P_{0,-1}$). 
\section{Symplectic form}\label{symplecticformWVG}
A coadjoint orbit $\mathcal{M}_\xi$ with $\xi\in\mathbf{g}^*$ is a symplectic submanifold of $\mathcal{M}$ which carries a natural symplectic structure $\omega_\xi$ denoted as Kirillov-Kostant-Souriau symplectic form \cite{Souriau,kostant,kirillov} defined by
\begin{align}\label{KKSform}
    \omega({X}_1(\xi),{X}_2(\xi))=-\xi([X_1,X_2])\,,
\end{align}
where $ X(\xi)$ is the vector field at $\xi\in\mathbf{g}^*$  generated by the coadjoint action of $X\in\mathbf{g}$; 
\begin{align}\label{coadjointalg}
&    \langle{X}(\xi),Y\rangle=\langle\text{ad}_X^*\xi,Y\rangle=-\langle\xi,[X,Y]\rangle\,.
\end{align}
Let us apply the definition \eqref{coadjointalg} to the coadjoint action of the warped Virasoro algebra $\mathbf{\hat g}$ on its dual space $\mathbf{\hat g}^*$. Using the vectors \eqref{adjvector} and covectors \eqref{covector} we have;
\begin{align}
{X}(\xi)=\text{ad}_{(\mathbf{v},\mathbf{a})}^*\left(\mathbf{b},\mathbf{c}\right)=(   \delta T(\tau)\extd\tau^2+\delta P (\tau)  \extd\tau,0)
\end{align}
where $\delta T$ and $\delta P$ are defined in \eqref{infinittranstran}.  

Now by having the knowledge of section \ref{coadjointsec} we are ready to calculate the value of the symplectic form $\omega$ \eqref{KKSform} for the warped Virasoro coadjoint orbit with constant representative at the point $(\mathbf{b},\mathbf{c})\in \mathbf{\hat g}^*$ on the pair of vector fields ${X}_1(\xi)$ and ${X}_2(\xi)$ on the orbit,
\begin{align}\label{symplectic12}
    \omega_{12}&=-\left\langle(\mathbf{b},\mathbf{c})\,,\big[(\mathbf{v}_1(\tau),\mathbf{a}_1)\,,(\mathbf{v}_2(\tau),\mathbf{a}_2)\big]\right\rangle\,\nn\\
    &=-\int_{S^1}\Big[T(\tau)\big(\epsilon_1(\tau)\epsilon_2'(\tau)-\epsilon_2(\tau)\epsilon_1'(\tau)\big)+P(\tau)\big(\epsilon_1(\tau)\sigma_2'(\tau)-\epsilon_2(\tau)\sigma_1'(\tau)\big)\nn\\&\qquad \;\qquad\quad +\frac{c_1}{12}\,\epsilon_1'(\tau)\epsilon_2''(\tau)-\frac{c_2}{4}\,\big(\epsilon_1'(\tau)\sigma_2'(\tau)-\epsilon_2'(\tau)\sigma_1'(\tau)\big)+\frac{c_3}{2}\sigma_1'(\tau)\sigma_2(\tau)\Big]
    \\
    &=-\frac{c}{12}\int_{S^1}\Big[\epsilon_1'\epsilon_2''-\big\{\tan\tfrac{\pi }{\beta}f;\tau\big\}\big(\epsilon_1\epsilon_2'-\epsilon_2\epsilon_1'\big)\Big]\nn\\
    &\quad+\kappa\int_{S^1}\Big[\epsilon_1'\sigma_2'-\epsilon_2'\sigma_1'+\big(\big(\tfrac{f''}{f'}+\tfrac{2\pi i}{\beta}f'\big)\tilde{g}'-\tilde{g}''\big)\big(\epsilon_1\epsilon_2'-\epsilon_2\epsilon_1'\big)+\big(\tfrac{f''}{f'}+\tfrac{2\pi i}{\beta}f'\big)\left(\epsilon_1\sigma_2'-\epsilon_2\sigma_1'\right)\Big]\nn\\
    &\quad-\frac{k}{2}\int_{S^1}\left[\sigma_1\sigma_2'+\tfrac12\tilde{g}'^2\big(\epsilon_1\epsilon_2'-\epsilon_2\epsilon_1'\big)+\tilde{g}'(\epsilon_1\sigma_2'-\epsilon_2\sigma_1')\right]\,,\label{symplectic122}
\end{align}
where in \eqref{symplectic12} we used the covector \eqref{covector} and the vector \eqref{adjvector} and used the pairing \eqref{pairing}. In  \eqref{symplectic122} the $\tau$-dependence of variables is implicit and for brevity we introduced $
\tilde{g}(\tau)=g(\tau)+\alpha f(\tau)
$. In transition from \eqref{symplectic12} to \eqref{symplectic122} we inserted the expression for $P(\tau)$ and $T(\tau)$ from \eqref{constrep} for the vacuum  orbit on the cylinder which is globally SL(2,$\mathbb R$)$\times$U(1) invariant.  Now we rephrase \eqref{symplectic122} in terms of differential forms;
\begin{align}
   \label{symplecticform} \omega
     &=-\frac{c}{24}\int_{S^1}\Big[\extd\epsilon'\wedge\extd\epsilon''-2\big\{\tan\tfrac{\pi }{\beta}f;\tau\big\}\extd\epsilon\wedge\extd\epsilon'\Big]\nn\\
     &\quad+\kappa\int_{S^1}    \big(\extd\epsilon'+\big(\tfrac{f''}{f'}+\tfrac{2\pi i}{\beta}f'\big)\extd\epsilon\big)\wedge\big(\extd\sigma+\tilde{g}'\extd\epsilon\big)'\nn\\
     &\quad-\frac{k}{4}\int_{S^1}    \big(\extd\sigma+\tilde{g}'\extd\epsilon\big)\wedge\big(\extd\sigma+\tilde{g}'\extd\epsilon\big)'\,.
     \end{align} 
Here $\extd$ acts on fields and not on coordinates and it essentially symbolizes a one-form on the coadjoint orbit. 
The symplectic form  in \eqref{symplecticform} is closed $\extd\omega=0$ by construction which can also be examined directly. It is invariant under the action of the twisted warped group \eqref{TWVgroup} and it is nondegenerate on the coadjoint orbit. In the following we find the finite form of the Symplectic 2-form.

According to the group operation \eqref{groupoper}  it is important to note that, $f$ as an element of the (twisted) warped Virasoro group \eqref{TWVgroup}, transforms under $\text{Diff}(S^1)$ while $\sigma$ transforms both under $\text{Diff}(S^1)$ and $C^\infty(S^1)$. So
we have;
\begin{align}\label{extdfextdg}
\extd f= f'\extd\epsilon\,,\qquad\qquad \extd g=\extd\sigma+g' \extd\epsilon\quad\text{with}\qquad g=\sigma\circ f\,.
\end{align}
Using \eqref{extdfextdg} one can rewrite \eqref{symplecticform} in a more abstract way as;
\begin{align}
   \label{symplecticformal} \omega
     &=-\frac{c}{24}\int_{S^1}\Big[\frac{\extd f'\wedge\extd f''}{f'^2}-\frac{4\pi^2}{\beta^2}\extd f\wedge\extd f'\Big]\nn\\&+\kappa\int_{S^1}    \extd\,\log(\exp \tfrac{2\pi i}{\beta}f)'\wedge\extd \tilde{g}'\nn\\&-\frac{k}{4}\int_{S^1}    \extd \tilde{g}\wedge\extd \tilde{g}'\,.
     \end{align} 
The first term in \eqref{symplecticformal} is the famous Kirillov-Kostant-Souriau symplectic form for the Virasoro-group \cite{Witten:1987ty,Alekseev:1988ce}  at central charge $c$ while the
last term  in \eqref{symplecticformal} is the contribution from the infinite dimensional Heisenberg group (U(1) Kac-Moody) at level $k/2$ \cite{Alekseev:1988ce,Wiegmann:1989hn} or the $k$-cocycle of the warped Virasoro group. 
The symplectic form of the twisted warped Virasoro group has an off-diagonal $\kappa$-contribution which is shown in the middle term which to the best of our knowledge is new. 
If $k\neq0$ we can diagonalize \eqref{symplecticformal}  and rewrite it in a more compact form
\begin{align}\label{sympcan}
    \omega
     =-\frac{c_{\text{\tiny eff}}}{24}\int_{S^1}\Big[\frac{\extd f'\wedge\extd f''}{f'^2}-\frac{4\pi^2}{\beta^2}\extd f\wedge\extd f'\Big]
     -\frac{k}{4}\int_{S^1}    \extd g_{\text{\tiny eff}}\wedge\extd g_{\text{\tiny eff}}'\,,
\end{align}
where 
\begin{align}\label{effectivecalpha}
      c_{\text{\tiny eff}}=c-\frac{24\kappa^2}{k}\,,\qquad 
     g_{\text{\tiny eff}}=g+\alpha_{\text{\tiny eff}}f -\frac{2\kappa}{k}\log f'\,,\qquad \alpha_{\text{\tiny eff}}=\alpha-\frac{2\kappa}{k}\Big(\frac{2\pi i }{\beta}\Big)\,.
\end{align}
It is essential to notice that at least classically when $k\neq0$, we could hide the $\kappa$-contribution from the vacuum orbit into the Virasoro and the Heisenberg orbits by doing shifts in \eqref{effectivecalpha} and the coadjoint orbit does not change in this case.  However the price one pays is that the field $g_{\text{\tiny eff}}$ unlike the field $\tilde g$ is no more pure imaginary and has a real part which is $\frac{2\kappa}{k}\log f'$. This field redefinition  will also happen at the level of the action \eqref{action2} and field equations \eqref{fequation}.
When the level $k$ is zero this change of variable is no more valid and we should directly work with \eqref{symplecticform} or \eqref{symplecticformal} at $k=0$.

\subsection{Pfaffian}

We may  evaluate the symplectic form on the coadjoint orbit around the identity element $f=\tau$ of Diff$(S^1)\ltimes C^\infty(S^1)$ such that $f(\tau)=\tau+\epsilon(\tau)$ and $g(\tau)=\sigma(\tau)$ by Fourier expanding the fluctuations as;
\begin{align}\label{fourierex}
   \extd \epsilon(\tau)=\frac{\beta}{2\pi}\sum_{n\in{\mathbb Z}} \extd\epsilon_n\,e^{-\frac{2\pi}{\beta} in\tau}\,,\qquad\qquad \extd \sigma(\tau)=\frac{\beta}{2\pi}\sum_{n\in{\mathbb Z}}\extd\sigma_{n}\,e^{-\frac{2\pi}{\beta} in\tau}\,. 
    \end{align}
In terms of these Fourier modes the symplectic form of the coadjoint orbit at identity becomes
\begin{align}\label{sympformmode1}
\omega&=\frac{\beta^2}{\pi i}\sum_{n\geq1}\big(\tfrac{c\pi^2}{6\beta^2}\,n^3+T^{\text{\scriptsize vac}}\,n\big)\extd\epsilon_n\extd\epsilon_{-n}
\nn\\
&+\frac{\beta^2}{2\pi i}\sum_{n\geq1}\big(\tfrac{2\pi i \kappa}{\beta}\, n^2+P^{\text{\scriptsize vac}}\,n\big)\extd\epsilon_n\extd{\sigma}_{-n}+\frac{\beta^2}{2\pi i}\sum_{n\geq1}\big(\tfrac{2\pi i \kappa}{\beta}\,n^2-P^{\text{\scriptsize vac}}\,n\big)\extd\epsilon_{-n}\extd{\sigma}_{n}\nn\\&+\frac{\beta^2}{2\pi i}\sum_{n\geq1}\tfrac{k}{2}\,n\extd{\sigma}_n\extd{\sigma}_{-n}\,.
\end{align}
We assume \begin{align}\label{holomorphicity}
\epsilon_n^*=\epsilon_{-n}\,,\qquad\qquad    \sigma_n^*=-\sigma_{-n}    \,,
\end{align}
 as a consequence of $f$ being essentially a real angle and $g$ being a pure imaginary function of $S^1$.
 We restricted the domain of sums in \eqref{sympformmode1} such that the symplictic form remains non-degenerate for generic values of couplings. Of course there are special values where the simplectic matrix becomes reducible and we should restrict it more. We will consider these cases separately below.
 
The natural volume form  (or measure of the integral) in an $M$-dimensional symplectic manifold $\mathcal M$ with coordinates $\xi^1,\cdots,\xi^M$ which is equipped with the antisymmetric symplectic matrix $\omega_{ij}$ is
\begin{align}
    \text{vol}_{\mathcal M}=\text{Pf}(\omega)\extd\xi^1\cdots\extd\xi^M\,.
\end{align}
In our case the matrix $\omega_{ij}$ has the following block form;
\begin{align}\label{symplecticmatrix}
\omega=
\left(
\begin{array}{c|c}
\omega_{\epsilon\epsilon}\; &\; \omega_{\epsilon\sigma} \\
\hline
\omega_{\epsilon\sigma}^{\text {\tiny T}} \;\;&\; \omega_{\sigma\sigma}
\end{array}
\right)\,.
\end{align}
The rank of matrices $\omega_{\epsilon\epsilon}$, $\omega_{\epsilon\sigma}$ and $\omega_{\sigma\sigma}$ are generically 
$2N$ 
 where $N$ is the dimensionality of the vector space which is a large number $\extd a_1\cdots\extd a_N$ and is the upper limit of sums in \eqref{sympformmode1}. In order to calculate the Pfaffian of the matrix \eqref{symplecticmatrix} we can factorize it as $\omega=B\Omega B^{\text{\tiny T}}$ with the non-singular matrix $B$ and use the following identity;
\begin{align}\label{identity}
    \text{Pf}(B\Omega B^{\text{\tiny T}})=\text{Pf}(\Omega)\text{det}(B)\,.
\end{align}
In our discussion this decomposition is made such that $B$ is upper/lower triangular. We keep $c$ being non-zero all the time, depending on other cocycles being zero or not we have the following three cases:

\begin{itemize}
\item In the case where $\kappa=0$ the symplectic form is reducible to
\begin{align}\label{sympformmode2}
\omega&=-\frac{c}{12}2\pi i\sum_{n\geq2}(n^3-n)\extd\epsilon_n\extd\epsilon_{-n}
+\frac{k}{2}\frac{\beta^2}{2\pi i}\sum_{n\geq1}n\extd\tilde{\sigma}_n\extd\tilde{\sigma}_{-n}
\end{align}
with $\extd\tilde{\sigma}_n=\extd\sigma_n+\alpha\extd\epsilon_n$. The Pfaffian of the antisymmetric symplectic matrix becomes
\begin{align}\label{Pfaffian1}
    \text{Pf}(\omega)&=\text{Pf}(\omega_{\epsilon\epsilon})\text{Pf}(\omega_{\sigma\sigma})\nn\\
    &=(-1)^{N-1}\prod_{n\geq2}^N\left(-\frac{c}{24}2\pi i(n^3-n)\right)\prod_{m\geq1}^N\left(-\frac{k}{4}\frac{\beta^2}{2\pi}im\right)\,.
\end{align}
  In this case $\omega_{\epsilon\epsilon}$ and $\omega_{\sigma\sigma}$ are $(2N-2)\times(2N-2)$ and $2N\times2N$ matrices and both should be non-degenerate in order to have a well-defined volume. This amounts to having the full SL$(2,{\mathbb R})\times$U(1) global identification on the orbit. In other words, in the absence of the  $\kappa$ cocycle the zero modes of the algebra which form the maximal finite subalgebra are $\epsilon_{0,\pm1}$ and $\sigma_0$ and they form the SL(2,$\mathbb R$)$\times$U(1).
\item In the case where $k=0$, the symplectic form is also reducible 
\begin{align}\label{sympformmode3}
\omega&=-\frac{c}{12}2\pi i\sum_{n\geq2}(n^3-n)\extd\epsilon_n\extd\epsilon_{-n}
\nn\\
&+\kappa\beta\sum_{n\geq2}(n^2-n)\extd\epsilon_n\extd\tilde{\sigma}_{-n}+\kappa\beta\sum_{n\geq1}(n^2+n)\extd\epsilon_{-n}\extd\tilde{\sigma}_{n}\,
\end{align}
and since the matrix $\omega_{\sigma\sigma}$ is identically zero the matrix $\omega_{\epsilon\epsilon}$ does not play any role in the Pfaffian of the matrix $\omega$; 
\begin{align}\label{Pfaffian2}
    \text{Pf}(\omega)&=(-1)^{N-1}\text{Pf}(\omega_{\epsilon\sigma})^2\nn\\
    &=(-1)^{N-1}\prod_{n\geq2}^N\frac{\kappa\beta}{2}(n^2-n)\prod_{m\geq1}^N\frac{\kappa\beta}{2}(m^2+m)\,.
\end{align}
In this case it is enough to have the $\omega_{\epsilon\sigma}$, which is a $(2N-1)\times(2N-1)$ matrix, being non-degenerate for having a well-defined symplectic volume in the phase space. As a consequence in the presence of $\kappa$ while $k=0$ the zero modes defining our orbit are $\epsilon_{0,1}$ and $\sigma_{-1,0}$ and they obey the centrally extended iso(1,1) algebra \cite{Afshar:2015wjm}.
    \item  In the most general case where all cocycles are non-zero we can find the Pfaffian using \eqref{sympformmode1}-\eqref{identity};
    \begin{align}
    \text{Pf}(\omega)&=\text{Pf}(\omega_{\epsilon\epsilon}+\omega_{\epsilon\sigma}\omega_{\sigma\sigma}^{-1}\omega_{\epsilon\sigma}^{\text{\tiny T}})\text{Pf}(\omega_{\sigma\sigma})\nn\\
    &=(-1)^{N-1}\prod_{n\geq2}^N\left(-\frac{c_{\text{\tiny{eff}}}}{24}2\pi i(n^3-n)\right)\prod_{m\geq1}^N\left(-\frac{k}{4}\frac{\beta^2}{2\pi}im\right)
\end{align}
where $c_{\text{\tiny{eff}}}=c-24\kappa^2/k$. Again we have restricted the domain of $n$ such that the determinant of the symplectic matrix is non-zero. In this case non-degeneracy of $\omega$ again leads to the SL$(2,{\mathbb R})\times$U(1) as the stabilizer of the orbit.%
\end{itemize}

This last result in the list is compatible with the diagonalization of the symplectic form in its finite form \eqref{sympcan}.

\section{Warped Schwarzian theory}\label{WSCHT}
In general the Hamiltonian corresponding to the imaginary time evolution $\tau\to\tau+\epsilon_0$ is  associated to the zero-mode of $T(\tau)$:
\begin{align}\label{action}
H=L_0=\int_{S^1} T(\tau)    \,.
\end{align}
Since we are considering the Euclidean theory the action is  equal to its imaginary-time Hamiltonian up to normalization. This is a consequence of applying the Duistermaat-Heckman formula which means the corresponding partition function in one-loop exact \cite{Stanford:2017thb}. This application is based on the fact that these theories arise from coadjoint orbits which are symplectic manifolds and possess a U(1) imaginary-time translation generator $L_0$. In fact this fact has been used to express the exponential of the Schwarzian action as the evolution operator in the quantum theory \cite{Mertens:2017mtv}. In the most general case we consider the following  one dimensional classical action;
\begin{align}\label{action}
S=\int\limits_0^\beta T(\tau)\,\extd\tau    =S^{(c)}+S^{(k)}+S^{(\kappa)}
\end{align}
where $S^{(c,k,\kappa)}$ correspond to different contributions of the three cocycles ($c,k,\kappa$) of the twisted warped Virasoro group respectively. 
Before going any further for simplicity we first make the following field redefinition;
\begin{align}\label{shiftedg}
\tilde{g}= g+\alpha\,f\,,
\end{align}
to eliminate the contribution of $\alpha$ terms from the action \eqref{action}. This however changes the periodicity property in $g$ as $\tilde{g}(\tau+\beta)=\tilde{g}(\tau)+\alpha\beta$. 
After we drop the total derivative contributions we have;
\begin{subequations}\label{ScSkSkap}
\begin{align}
    S^{(c)}&=-\frac{c}{12}\int\limits_0^\beta \left\{\tan\tfrac{\pi }{\beta}f;\tau\right\}\,d\tau\,,\label{actionc}\\
    S^{(k)}&=\frac{k}{4}\int\limits_0^\beta \tilde{g}'^2\extd\tau\,,\\
    S^{(\kappa)}&=-\kappa\int\limits_0^\beta\frac{ \big(\exp \frac{2\pi i}{\beta}f\big)''}{ \big(\exp\frac{2\pi i}{\beta} f\big)'}\,\tilde{g}'\extd\tau\,.\label{actionkap}
\end{align}
\end{subequations}
If $k\neq0$ we can always absorb the contribution of $ S^{(\kappa)}$ into $S^{(c)}$ and $S^{(k)}$ up to total derivative terms, leading to the following action;
\begin{align}\label{action2}
  S=-\frac{c_{\text{\tiny eff}}}{12}\int\limits_0^\beta \left\{\tan\tfrac{\pi }{\beta}f;\tau\right\}+\frac{k}{4}\int\limits_0^\beta {g}_{\text{\tiny eff}}'^2 \,,
\end{align}
with effective values defined in \eqref{effectivecalpha}.
As obviously seen from \eqref{action2} the action for the non-vanishing $U(1)$ level corresponds to  the Schwarzian action of the vacuum Virasoro coadjoint orbit as diffeomorphisms of the circle with an improvement in the central charge decoupled from a free scalar ${g}_{\text{\tiny eff}}$. We expect that the measure of the integration remains invariant under these shifts in $g$;
\begin{align}\label{gshift}
    g\to {g}_{\text{\tiny eff}}= g+\alpha f-\frac{2\kappa}{k}\left(\log f'+\frac{2\pi i}{\beta}f\right)\,,
\end{align}
as it can be considered as part of the group action on $g$. However the periodicity property of $g$ will be twisted as a consequence of multiple shifts;
\begin{align}\label{boundarycondg}
    g(\tau+\beta)=g(\tau)+\alpha_{\text{\tiny eff}}\,\beta\,.
\end{align}
The  action \eqref{action2} when $\kappa=0$ has been studied extensively as the effective action of the complex SYK model which has a U(1) global symmetry.\footnote{For some studies on the SYK model with global symmetries see \cite{PhysRevLett.70.3339,Fu:2016vas,Gross:2016kjj,Bhattacharya:2018nrw,Bulycheva:2017uqj,Yoon:2017nig,Narayan:2017hvh,Grumiller:2017qao,Gonzalez:2018enk,Liu:2019niv}.} The goal in this note is to consider the consequences when $\kappa\neq0$. Specially once $k=0$  the shift \eqref{gshift} is no more valid. 
\subsection{Saddle points of the action}
Upon variation of the the action \eqref{action} w.r.t. $f$ and $\tilde g$ \eqref{shiftedg}, in general we have the following field equations;
\begin{subequations}\label{eoms}
\begin{align}\label{eom1}
\delta f\,,&\qquad \frac{c}{12}\left[\frac{4\pi^2}{\beta^2}f'+\frac{1}{f'}\left(\frac{f''}{f'}\right)'\right]'+\kappa\left[\frac{{\tilde g}''}{f'}-\frac{2\pi i}{\beta}{\tilde g}'\right]'    =0\,,\\
\delta {\tilde g}\,,&\qquad \kappa\left[\frac{f''}{f'}+\frac{2\pi i}{\beta}f'\right]'-\frac{k}{2}{\tilde g}''=0\,.\label{eom2}
\end{align}
\end{subequations}
In general when $\kappa\neq0$ these two equations are coupled. If $k\neq0$, one can solve the  second equation for ${\tilde g}'$  and insert it into the first equation to obtain the $f$-equation; 
\begin{align}\label{fequation}
\frac{c_{\text{\tiny eff}}}{12}\frac{1}{f'}\bigg\{\tan\frac{\pi f }{\beta};\tau\bigg\}' =0\,.
\end{align}
Without loss of generality we assume $c_{\text{\tiny eff}}\neq0$, otherwise the function $f(\tau)$ remains arbitrary and the classical action becomes zero. Furthermore, $f'>0$ and finite as a consequence of $f(\tau)$ being  an orientation-preserving diffeomorphism. Thus the equation \eqref{fequation} can be solved as
\begin{align}\label{fsolution}
   \tan\tfrac{\pi }{\beta}f(\tau)=\frac{a/\lambda\sin{\lambda\tau}+b\cos{\lambda\tau}}{c/\lambda\sin{\lambda\tau}+d\cos{\lambda\tau}}\,.
\end{align}
The numerator and the denominator in \eqref{fsolution} have to be linearly independent so we should have $ad-bc\neq0$. The solution to the second equation \eqref{eom2} determines the field ${\tilde g}(\tau)$;
\begin{align}\label{gsolution}
    \tilde{g}(\tau)&=\tfrac{2\kappa}{k}\log [\exp\tfrac{2\pi i}{\beta}f]'+B_0\tau+C_0\,.
\end{align}
At finite temperature $\beta$ and finite chemical potential $\alpha$, the periodicity properties \eqref{boundarycond} and \eqref{boundarycondg} fix  integration constants  as $\lambda=\tfrac{\pi n}{\beta}$  and $B_0=\alpha_{\text{\tiny eff}}$. The zero temperature limit ($\beta\to\infty$) of these solutions is taken by replacing $\tan\tfrac{\pi }{\beta}f$ and $\exp\tfrac{2\pi i}{\beta}f$ in \eqref{fsolution} and \eqref{gsolution}  by $f$  while in this case $\lambda$ is not quantized anymore.

\subsection{On-shell action}\label{onshell}
The corresponding on-shell value of the action for the classical solution \eqref{fsolution}-\eqref{gsolution} is;
\begin{align}\label{onshell1}
   F\supset \beta^{-1} S_{_{\text{\tiny on-shell}}}=-\frac{c_{\text{\tiny eff}}}{6}\lambda^2+\frac{k}{4}\alpha_{\text{\tiny eff}}^2\,,
    \end{align}
where the effective values $c_{\text{\tiny eff}}$ and $\alpha_{\text{\tiny eff}}$ are given in \eqref{effectivecalpha}. This sounds similar to the finite temperature contribution to the classical free energy of the low energy complex SYK by replacing the bare values of the central charge and the chemical potential with physical ones as in \eqref{effectivecalpha}. However there is an essential difference, namely the effective chemical potential  depends on  $\beta$. This would change the behaviour of the density of states.

In the case where $k=0$ the solution \eqref{fsolution}-\eqref{gsolution} does not hold.  In this case one can first solve $f(\tau)$ from  \eqref{eom2} and plug the solution into equation \eqref{eom1} and solve for ${\tilde g}(\tau)$. In terms of the new variable $\exp\tfrac{2\pi i}{\beta}f$, from equation \eqref{eom2} we have;
 \begin{align}
    (\exp\tfrac{2\pi i}{\beta}f)''=2i\lambda(\exp\tfrac{2\pi i}{\beta}f)'\,,\qquad\qquad\lambda=\frac{\pi n}{\beta}\,,\label{scalesym}
 \end{align}
 whereas equation \eqref{eom1} can be rewritten as; 
\begin{align}\label{eoms2}
\frac{c}{12}\frac{\big\{\exp\frac{2\pi i}{\beta}f;\tau\big\}'}{(\tan\frac{\pi}{\beta}f)'} -\frac{2\pi i\kappa}{\beta}\left[\frac{({\tilde g}' \exp\tfrac{-2\pi i}{\beta}f)'}{( \exp\tfrac{-2\pi i}{\beta}f)'}\right]'    =0\,.
\end{align}
The equation \eqref{scalesym} shows that taking derivatives only scales the classical solutions. As a consequence $\big\{\exp\frac{2\pi i}{\beta}f;\tau\big\}=2\lambda^2$ and the contribution of the term with derivative of the Schwarzian to the equation \eqref{eoms2} is zero. In this case ($k=0$) 
the on-shell action after using \eqref{scalesym} is very easy to find from \eqref{actionkap};
\begin{align}\label{onshell2}
    F\supset\beta^{-1} S_{_{\text{\tiny on-shell}}}
    =-\frac{c}{6}\lambda^2-2i\alpha\kappa\,\lambda \,.
\end{align}
The exact finite temperature solutions of both $f$ and $g$ in this case are as follows;
\begin{align}
    \exp{\tfrac{ 2 \pi i }{\beta}f(\tau)}= C_0e^{2i\lambda\tau}+C_1\,,\qquad \tilde{g}(\tau)=C_2\exp{\tfrac{ 2 \pi i }{\beta}f(\tau)}+B_0\tau+C_3
\end{align}
where $C_i$'s are some arbitrary constants and $B_0=\alpha=-i\mu$ accounts for the chemical potential. Again the zero temperature solutions can be obtained by replacing $\exp\tfrac{ 2 \pi i }{\beta}f$ with $f$ and considering $\lambda$ as being unquantized.

\section{Partition function}\label{pathintegral}
In this section we set to evaluate the partition function for the warped Schwarzian theory defined by the Euclidean action \eqref{action} as the integral
\begin{align}\label{partitionfun}
    Z=\int_{\mathcal M} {\mathcal D}f {\mathcal D}g \,e^{-S[f,g]}\,.
\end{align}
The fields $f$ and $g$ subject to their boundary conditions
are elements of the constant representative coadjoint orbits of the twisted warped Virasoro group. The phase space which restricts the integration space in \eqref{partitionfun} is  the infinite dimensional quotient space
\begin{align}\label{coadjointman}
    {\mathcal M}=\frac{\text{Diff}(S^1)\ltimes C^\infty(S^1)}{\text{SL}(2,\mathbb{R})\times \text{U}(1)}\,.
\end{align}
The coadjoint orbit \eqref{coadjointman} as a manifold is symplectic (and in this case also K\"ahler) and as a
consequence the measure of integration in \eqref{pathintegral} is fixed using the Pfaffian of the symplectic matrix. The Duistermaat-Heckman (DH) theorem, applied to symplectic manifolds endowed with a U(1) time translation generators $L_0$ playing the role of the action in the integral \eqref{partitionfun}, states that the path integral is 1-loop exact;
\begin{align}
    Z=e^{-S^{(0)}}
     Z_{\text{\tiny 1-loop}}\,.
\end{align}
In order to evaluate the one-loop integral it is enough to expand around the classical solution to the equations of motion. Around the saddle point we can expand as $f(\tau)\simeq\tau +\epsilon(\tau)$ and since $g=\sigma\circ f$, we have $g(\tau)\simeq\sigma(\tau)+\epsilon(\tau)\sigma'(\tau)$. We can think of $\epsilon(\tau)$ and $\sigma(\tau)$ as Goldstone modes for the broken SL(2,${\mathbb R}$) and U(1).
The action \eqref{action} to quadratic order in $\epsilon(\tau)$ and $\sigma(\tau)$ becomes $S=S^{(0)}+S^{(2)}$ with;
\begin{eqnarray}
\label{S0}
S^{(0)}&=&\beta T^{\text{\scriptsize vac}}=-\frac{c\pi^2}{6\beta}-2\pi i\alpha\kappa+\frac{k\beta}{4}\alpha^2\,,\\
S^{(2)}&=&\int\limits_0^\beta\Big[\frac{c}{24}\Big(\epsilon''^2-\big(\tfrac{2\pi}{\beta}\big)^2\epsilon'^2\Big)+\frac{k}{4}\big(\sigma'+\alpha\epsilon'\big)^2-\kappa\Big(\epsilon''+\tfrac{2\pi i}{\beta}\epsilon'\Big)\big(\sigma'+\alpha\epsilon'\big)\Big]\extd\tau\,.\label{S1kneqz}
\end{eqnarray}
The expression for $S^{(0)}$ is the saddle point contribution 
$S_{\text{\tiny on-shell}}$ in \eqref{onshell1} for $\lambda=\pi/\beta$ as expected. 
The goal would be to compute the Euclidean one-loop path integral 
\begin{align}\label{1looppartitionfun}
     Z_{\text{\tiny 1-loop}}=\int_{\mathcal M} {\mathcal D}\epsilon\, {\mathcal D}\tilde{\sigma} \,e^{-S^{(2)}[\epsilon,\tilde{\sigma}]}\,.
\end{align}
Here ${\tilde \sigma}=\sigma'+\alpha\epsilon'$. This change of variable leaves the measure of the integral invariant as $\sigma\to{\tilde \sigma}$ is a group action.  In order to perform the path integral on the quotient space we should not include zero modes or the kernel of the symplectic from  \eqref{sympformmode1} in the measure of integration. 
The periodicity properties of $f$ and $g$ in \eqref{boundarycond} and \eqref{boundarycondg} imply that $\epsilon(\tau+\beta)=\epsilon(\tau)$ and $\sigma(\tau+\beta)=\sigma(\tau)$. 
We employ the Fourier mode expansion  of the fluctuations;
\begin{align}\label{fourierex}
   \epsilon(\tau)=\frac{\beta}{2\pi}\sum_{n\in{\mathbb Z}} \epsilon_n\,e^{-\frac{2\pi}{\beta} in\tau}\,,\qquad\qquad {\sigma}(\tau)=\frac{\beta}{2\pi}\sum_{n\in{\mathbb Z}} {\sigma}_n\,e^{-\frac{2\pi}{\beta} in\tau}\,,
    \end{align}
and evaluate the quadratic effective action \eqref{S1kneqz} to quadratic order in fluctuations around the SL(2,$\mathbb R$)$\times$U(1) vacuum.
We find the individual contribution to each cocycle as follows;
\begin{align}\label{actionmodes}
S^{(2)}&= 
\frac{c\pi^2}{3\beta}\sum_{n\geq2}n^2(n^2-1)|\epsilon_n|^2\nn\\
&+2\pi i\kappa\sum_{n\geq2}n(n^2-n)\epsilon_n\tilde{\sigma}_{-n}+2\pi i\kappa\sum_{n\geq1}n(n^2+n)\epsilon_{-n}\tilde{\sigma}_{n}\nn\\
&+ \frac{k\beta}{2}\sum_{n\geq1}n^2\left|\tilde{\sigma}_n\right|^2\,,
\end{align}
where $\tilde{\sigma}_n=\sigma_n+\alpha\epsilon_n$.
Using the symplictic from \eqref{sympformmode1} and the corresponding Pfaffian we can perform the 1-loop path integral \eqref{1looppartitionfun}. 
We find it useful to discuss cases where the twist term $\kappa$ or the level $k$ is zero separately.

\subsection{Untwisted warped Schwarzian theory with non-zero level}\label{untwistedleveled}
The path integral of the twistless case when $\kappa=0$ is the same as for the complex SYK model in the low energy regime carried out in \cite{Stanford:2017thb,Mertens:2019tcm,Liu:2019niv}. The modes $\epsilon$ and $\tilde\sigma$ are decoupled in  both the action and the measure of the path integral and we can evaluate the 1-loop path integral by evaluating each contribution separately.
Using the Pfaffian \eqref{Pfaffian1}, the contribution of each  piece is as follows;\footnote{We have $\extd^2\epsilon_n=\extd\epsilon_n^R\extd\epsilon_n^I=\frac{i}{2}\extd\epsilon_n\extd\epsilon_{-n}$ and $\extd^2\sigma_n=\extd\sigma_n^R\extd\sigma_n^I=-\frac{i}{2}\extd\sigma_n\extd\sigma_{-n}$. 
}
\begin{align}\label{1loopZcZk}
Z_{\text{\tiny 1-loop}}^{(c)}&=(-1)^{\frac{N(N-1)}{2}} \prod_{n\geq2}\frac{c\pi}{6}(n^3-n)\int\extd^2\epsilon_n\exp\Big( -\frac{c\pi^2}{3\beta}n^2(n^2-1)|\epsilon_n|^2\Big)\\
&=\pm2\beta^{-1}\prod_{n\geq1}\frac{\beta}{2n}=\pm4\beta^{-1}\sqrt{\frac{\pi}{\beta}}\,,\\
Z_{\text{\tiny 1-loop}}^{(k)}&=(-1)^{\frac{N(N-1)}{2}} \prod_{n\geq1}\frac{k\beta^2}{4\pi}n\int\extd^2\tilde{\sigma}_n\exp\Big(- \frac{k\beta}{2}n^2\left|\tilde{\sigma}_n\right|^2\Big)\\
&=\pm\prod_{n\geq1}\frac{\beta}{2n}=\pm2\sqrt{\frac{\pi}{\beta}}\,.
\end{align}
The Gaussian integrals can easily be evaluated. In order to evaluate the infinite products we used the zeta function regularization 
\begin{align}
    \prod_{n=1}\frac{\beta}{2n}=\exp\{-\frac{\extd}{\extd s}\sum_{n=1}(\tfrac{\beta}{2n})^{-s}\Big|_{s=0}\}=\exp\{(\log\tfrac\beta2)\zeta(0)-\zeta'(0)\}=2\sqrt{\frac\pi\beta}\,.
\end{align}  The final answer is the product of contributions from the $c$- and the $k$-cocycles; 
\begin{align}
Z_{\text{\tiny 1-loop}}^{(\kappa=0)}=Z_{\text{\tiny 1-loop}}^{(c)}Z_{\text{\tiny 1-loop}}^{(k)}\propto\frac{1}{\beta^2}\,.
\end{align}
The proportionality constant is state independent and irrelevant. This one-loop result matches with our expectation regarding the number of zero modes which is four in the case of warped Virasoro algebra. 
The full partition function in this case is
\begin{align}
    Z(\beta,\alpha)\propto\frac{1}{\beta^2}\exp\Big(\frac{c\pi^2}{6\beta}-\frac{k\beta}{4}\alpha^2\Big)\,.
\end{align}

\subsection{Twisted warped Schwarzian theory at level zero}
Another interesting case would be to carry out the partition function when the U(1) level is vanishing $k=0$ but $\kappa\neq0$. At the first glance to the action \eqref{action}, in this case the $g$ field plays the role of a Lagrange multiplier that can be integrated out;
 \begin{align}
     Z&=\int_{\mathcal{ M}_{(k=0)}} {\mathcal D}f \,\delta\big[\big(\log f'+\tfrac{2\pi i}{\beta}f\big)''\big] \,e^{-S_{\text{\tiny Sch}}[f]}\,.
 \end{align}
 The path integral would then reduce to evaluating a Dirac delta functional in the integral. This is again naive as one should be cautious with the domain of the integration. In the  path integral \eqref{partitionfun}, we should avoid integrating over zero modes that are globally quotiented in the integration space \eqref{coadjointman}. Again using the mode expansion of the action \eqref{actionmodes} and the Pfaffian of the symplectic matrix \eqref{Pfaffian2} in this case we have
\begin{align}
    Z_{\text{\tiny 1-loop}}^{(k=0)}= &(-1)^{N-1}\prod_{n\geq2}\frac{\kappa\beta}{2}(n^2-n)\int\extd\epsilon_n\extd\tilde{\sigma}_{-n}\exp\left(-\frac{c\pi^2}{6\beta}n^2(n^2-1)|\epsilon_n|^2-{2\pi i\kappa}\,n(n^2-n)\epsilon_n\tilde{\sigma}_{-n}\right)\nn\\&\prod_{m\geq2}\frac{\kappa\beta}{2}(m^2+m)\int\extd\epsilon_{-m}\extd\tilde{\sigma}_{m}\exp\left(-\frac{c\pi^2}{6\beta}m^2(m^2-1)|\epsilon_m|^2-{2\pi i\kappa}\,m(m^2+m)\epsilon_{-m}\tilde{\sigma}_{m}\right)\nn\\
     &\kappa\beta\int\extd\epsilon_{-1}\extd\tilde{\sigma}_1 \exp\left(-{4\pi i\kappa}\,\epsilon_{-1}\tilde{\sigma}_{1}\right)\nn\\
     =&(-1)^{N-1}\,I_{1,-1}\,\prod_{n\geq2}\kappa^2\beta^2n^2(n^2-1)\int\extd^2\epsilon_n\extd^2\tilde{\sigma}_{n}\exp\left(-A_n|\epsilon_n|^2-\epsilon_n^RJ_n-\epsilon_n^IK_n\right)
    \nn\\    
         =&(-1)^{N-1}\,I_{1,-1}\,\prod_{n\geq2}\frac{3\kappa^2\beta^3}{c\pi}\int\extd^2\tilde{\sigma}_{n}\exp\left(\frac{12\kappa^2\beta}{c}n^2|\sigma_n|^2\right)
    \nn\\
     =&I_{1,-1}\,\prod_{n\geq2}\frac{\beta^2}{4n^2} 
    =\frac\beta2\frac{16\pi}{\beta^3}
     = \frac{8\pi}{\beta^2}
\end{align}
where
\begin{align}
A_n&=\frac{c\pi^2}{3\beta}n^2(n^2-1)\,,\quad J_n=4\pi \kappa\left(in^2\sigma_n^R-n^3\sigma_n^I\right)\,,\quad K_n=4\pi \kappa\left(n^3\sigma_n^R+in^2\sigma_n^I\right)\,.
 \end{align}
We evaluated the norm of the integral $I_{1,-1}$ by the doubling trick\footnote{Since we assumed $\sigma$ to be pure imaginary we assure that $\kappa$ is real.};
\begin{align}
|I_{1,-1}|^2&=    -4\kappa^2\beta^2\int\extd^2\epsilon_{1}\extd^2\tilde{\sigma}_1 \exp\left(-8\pi \kappa\left[\epsilon^I_{1}\tilde{\sigma}_1^R-\epsilon^R_{1}{\tilde\sigma}_1^I\right]\right)=\frac{\beta^2}{4}\,.
\end{align}
We can also do a direct calculation by expanding $\epsilon_{-1}$ and $\sigma_1$;
\begin{align}
    I_{1,-1}&=
        \frac{\kappa\beta}{4}\Big(\iint\limits_{-\infty}^\infty\extd\epsilon_{-1}^R\extd\tilde{\sigma}_{1}^R-\iint\limits_{-\infty}^\infty\extd\epsilon_{-1}^I\extd\tilde{\sigma}_{1}^I+i\iint\limits_{-\infty}^\infty\extd\epsilon^R_{-1}\extd{\tilde\sigma}_1^I+i\iint\limits_{-\infty}^\infty\extd\epsilon^I_{-1}\extd\tilde{\sigma}_1^R\Big)\nn\\ &\quad\qquad\cdot\exp\left(4\pi \kappa\left[\epsilon^R_{-1}{\tilde\sigma}_1^I+\epsilon^I_{-1}\tilde{\sigma}_1^R\right]-4\pi i\kappa[\epsilon_{-1}^R\tilde{\sigma}_{1}^R-\epsilon_{-1}^I\tilde{\sigma}_{1}^I]\right)=-\frac\beta2\,.
\end{align}
\subsection{Twisted warped Schwarzian theory with non-zero level}
Once $\kappa\neq0$ and $k\neq0$, it is possible to hide the $\kappa$-contribution to the quadratic action in $c$- and  $k$-terms
\begin{align}\label{action42}
S^{(2)}&=\int\limits_0^\beta\Big[\frac{c_{\text{\tiny eff}}}{24}\Big(\epsilon''^2-\big(\tfrac{2\pi}{\beta}\big)^2\epsilon'^2\Big)+\frac{k}{4}\sigma'^2
_{\text{\tiny eff}}\Big]\extd\tau\,.
\end{align}
where
\begin{align}\label{sigmaeff}
    \sigma_{\text{\tiny eff}}=\sigma+\alpha_{\text{\tiny eff}}\epsilon-\tfrac{2\kappa}{k}\epsilon'
\end{align} with the effective values $c_{\text{\tiny eff}}$ and $\alpha_{\text{\tiny eff}}$ being defined in \eqref{effectivecalpha}. A quick conclusion would be to map the problem to the previous case \ref{untwistedleveled} where $\kappa=0$ and $k\neq0$ with physical quantities, $c$ and $\alpha$ being modified to effective ones \eqref{effectivecalpha}. 
However we note that the shift in $\tilde\sigma$ in \eqref{sigmaeff} is violating the holomorphicity of the $\sigma$ field. In other words, the new $\sigma_{\text{\tiny eff}}$ is a complex field whose real and pure imaginary values are independent from each other. This means that the positive and negative Fourier modes $(\sigma_{\text{\tiny eff}})_n$ are no more related to each other as they were before \eqref{holomorphicity}. 
So in order to keep track of all terms we preferably work with old variables $\epsilon$ and $\tilde\sigma$ in the action;
\begin{align}
    Z_{\text{\tiny 1-loop}} 
     =&K_{1,-1}\,\prod_{n\geq2}\frac{c_{\text{\tiny eff}}}{12}\frac{k\beta^2}{2}n(n^3-n)\int\extd^2\epsilon_n\extd^2\tilde{\sigma}_n\exp\left(-A_n|\epsilon_n|^2-\frac{k\beta}{2}n^2|{\tilde\sigma}_n|^2-\epsilon_n^RJ_n-\epsilon_n^IK_n\right)
       \nn\\    
         =&K_{1,-1}\,\prod_{n\geq2}\frac{c_{\text{\tiny eff}}k\beta^3}{c\,8\pi}\int\extd^2\tilde{\sigma}_{n}\exp\Big(-\frac{k\beta}{2c}\big[c-24\kappa^2/k\big]n^2|\sigma_n|^2\Big)
    \nn\\
     =&K_{1,-1}\,\prod_{n\geq2}\frac{\beta^2}{4n^2} 
    \sim\frac{8\pi}{\beta^2}\,.
\end{align}
In this case we used;
\begin{align}
     K_{1,-1}&=\frac{k\beta^2}{4\pi}\int
     \extd^2\tilde{\sigma}_1 \exp\left(-\frac{k\beta}{2}|{\tilde\sigma}_1|^2-4\pi i\kappa\,\epsilon_{1}^*\tilde{\sigma}_{1}\right)=\frac{\beta}{2}\,.
\end{align}
The full partition function of the warped Schwarzian theory is
\begin{align}\label{WSChpartition}
    Z(\beta,\alpha)\propto\frac{1}{\beta^2}\exp\Big(\frac{c\pi^2}{6\beta}+2\pi i\alpha\kappa-\frac{k\beta}{4}\alpha^2\Big)\,,
\end{align}
where $\alpha=-i\mu$ with $\mu$ playing the role of a chemical potential to the U(1) charge.

In order to compare the spectrum of the warped Schwarzian theory with its Schwarzian parent we separate the full partition function \eqref{WSChpartition} as follows
\begin{align}\label{WSChpartition}
    Z_{\text{\tiny WSch}}(\beta,\mu)=\frac{Z_{\text{\tiny Sch}}(\beta)}{\sqrt\beta}\exp\Big(2\pi \mu\kappa+\frac{k\mu^2}{4}\beta\Big)\,.
\end{align}
The density of states can be determined as the inverse Laplace transform to the partition function.
If we assume $k<0$, and shift the zero point energy to $-\tfrac{k\mu^2}{4}$ we have
\begin{align}\label{densityofstates}
    \rho_{\text{\tiny WSch}}(E-\tfrac{k\mu^2}{4},\mu)&=\frac{e^{2\pi \mu\kappa}}{\sqrt\pi}\,\int\limits_0^{E}\frac{\rho_{\text{\tiny Sch}}(E')}{\sqrt{E-E'}}\extd{ E'} \\
      &=2\,e^{2\pi \mu\kappa}\sqrt{6 \,E\,/c}\,I_1\big(2\sqrt{c\,E\,/6}\big)\,,\qquad E>0\,,
\end{align}
where $\rho_{\text{\tiny Sch}}(E)=\frac{2\sqrt{6}}{\sqrt{c\pi}}\sinh\big(2\pi\sqrt{c\,E\,/6}\,\big)$ and $I_1(x)=-iJ_1(ix)$ is the modified Bessel function of the first kind. Asymptotically for $cE\gg1$ the density of states as a function of $E$ grows like $\rho\sim\frac{e^{2\sqrt{c\,E\,/6}}}{E^{1/4}}$ while for small values above the zero point energy $ cE\ll1$ it behaves linearly $\rho\sim E$. The result in \eqref{densityofstates} also holds at the critical point when $k=0$. The $k>0$ does not lead to a well defined density, however if $\kappa=0$ we can have $\alpha$ being real ($\mu$ pure imaginary) and we are back to the above case.

\section{Thermodynamics}\label{discussion}
In this section we discuss the thermodynamics of the warped Schwarzian theory \eqref{action}. When the $\kappa$-term in \eqref{action} is zero, this theory coincides with the low-energy effective action for the complex SYK model at large $N$. The fluctuations of energy above the ground state $\mathcal{E}$ and the total charge $\mathcal{Q}$ associated to the U(1) symmetry can be obtained from \eqref{S1kneqz} as \cite{Davison:2016ngz},
\begin{align}\label{EQ}
\mathcal{E}(\tau)-\mu\mathcal{Q(\tau)}=\frac{\delta S^{(2)}}{\delta\epsilon'(\tau)}\,,\qquad \mathcal{Q}(\tau)=i\frac{\delta S^{(2)}}{\delta\sigma'(\tau)}\,.
\end{align}
We have,
\begin{subequations}
\begin{align}
       \mathcal{E}(\tau)&=-\frac{c}{12}\Big[\epsilon'''+\left(\tfrac{2\pi}{\beta}\right)^2\epsilon'\Big]+\kappa\Big[\tilde{\sigma}''-\tfrac{2\pi i}{\beta}\tilde{\sigma}'\Big]\,,\\
    \mathcal{Q}(\tau)&=\frac{ik}{2}\tilde{\sigma}'-i\kappa\Big[\epsilon''+\tfrac{2\pi i}{\beta}\epsilon'\Big]=\frac{ik}{2}{\sigma}'_{\text{\tiny eff}}\,.
\end{align}
\end{subequations}
 When $k\neq0$,\footnote{Here we consider the case where $k$ is non-vanishing, the case where $k=0$ has been considered in \cite{Afshar:2019axx}.
} in order to compute the corresponding two-point correlators among  $\mathcal{E}$ and $\mathcal{Q}$ in the warped Schwarzian theory, we first exploit the action presented in \eqref{action42} and its analogy to the complex SYK model to present the correlators among $\epsilon$ and $\sigma_{\text{\tiny eff}}$ \cite{Maldacena:2016hyu,Davison:2016ngz};
 \begin{align}
     &\langle\epsilon(\tau)\epsilon(0)\rangle=\frac{3\beta^3}{4\pi^4c_{\text{\tiny eff}}}\left[\frac{\pi^2}{6}+1-\frac12\big(\tfrac{2\pi\tau}\beta-\pi\big)^2+\frac{5}{2}\cos\big(\tfrac{2\pi\tau}{\beta}\big)+\big(\tfrac{2\pi\tau}\beta-\pi\big)\sin\big(\tfrac{2\pi\tau}{\beta}\big)\right]\,,\\
     &\langle{\sigma}_{\text{\tiny eff}}(\tau){\sigma}_{\text{\tiny eff}}(0)\rangle=\frac{\beta}{2\pi^2k}\left[\frac12\big(\tfrac{2\pi\tau}\beta-\pi\big)^2-\frac{\pi^2}{6}\right]\,,
 \end{align}
where $\sigma_{\text{\tiny eff}}=\tilde{\sigma}-\frac{2\kappa}{k}(\epsilon'+\frac{2\pi i}{\beta}\epsilon)$ as given in \eqref{sigmaeff}. Furthermore, using the above correlators and the fact that $\langle\epsilon(\tau)\sigma_{\text{\tiny eff}}(0)\rangle=0$ we can obtain the two-point correlators for $\tilde{\sigma}$;
\begin{subequations}
\begin{align}
    &\langle\tilde{\sigma}(\tau)\epsilon(0)\rangle=\frac{2\kappa}{k}\left(\partial_\tau+\tfrac{2\pi i}{\beta}\right)\langle\epsilon(\tau)\epsilon(0)\rangle\,,\\
    &\langle\epsilon(\tau)\tilde{\sigma}(0)\rangle=-\frac{2\kappa}{k}\left(\partial_\tau-\tfrac{2\pi i}{\beta}\right)\langle\epsilon(\tau)\epsilon(0)\rangle\,,\\
    &\langle\tilde{\sigma}(\tau)\tilde{\sigma}(0)\rangle=\langle\sigma_{\text{\tiny eff}}(\tau)\sigma_{\text{\tiny eff}}(0)\rangle-\frac{4\kappa^2}{k^2}\left(\partial^2_\tau+\big(\tfrac{2\pi }{\beta}\big)^2\right)\langle\epsilon(\tau)\epsilon(0)\rangle\,.
\end{align}
\end{subequations}
Using these results, we can compute the two-point correlators between the conserved quantities in \eqref{EQ} which turn out to be $\tau$-independent;
\begin{align}\label{correlators}
    \langle \mathcal{E}(\tau) \mathcal{E}(0)\rangle=\frac{\pi ^2 c}{3 \beta ^3}\,,\qquad
    \langle \mathcal{E}(\tau) \mathcal{Q}(0)\rangle=-\frac{2 \pi   \kappa  }{\beta ^2 }\,,\qquad
    \langle \mathcal{Q}(\tau) \mathcal{Q}(0)\rangle=\frac{k}{2\beta}\,.
\end{align}
Obviously when $\kappa\neq0$ we see a new non-zero correlator in \eqref{correlators} in comparison to the complex SYK model. This explicitly demonstrates the role of all three cocycles of the warped Virasoro group to the complex SYK model. For comparison with reported results in \cite{Davison:2016ngz}, one can rename $c$ and $k$ in \eqref{correlators} in terms of the phenomenological couplings $\gamma$ (heat capacity) and $K$ (zero-temperature compressibility) in the complex SYK model as follows;
\begin{align}
    c\to\frac{3\gamma N}{\pi^2}\,,\qquad\qquad k\to 2NK\,.
\end{align}

\acknowledgments
The author thanks St\'ephane Detournay, Daniel Grumiller and Blaza Oblak for collaboration and discussions on aspects of the warped Virasoro group. He also thanks 
 Seyyed Mohammad Hassan Halataei, Ali Mollabashi,  Friedrich Sch\"oller and Raphaela Wutte
   for discussions. He was supported by the HEE-project P-28751. He also acknowledges the Iran-Austria IMPULSE project grant supported and run by Kharazmi University and OeAD-GmbH. He thanks the  Erwin Schr\"odinger International Institute for Mathematics and Physics (ESI) for hospitality during the Workshop  `Higher spins  and  holography'  in  March/April  2019.


\bibliographystyle{fullsort}
\bibliography{references}

\end{document}